\documentstyle[12pt,axodraw]{article}

\catcode`@=12
\newcommand{\beq}{\begin{equation}}
\newcommand{\eeq}{\end{equation}}

\newcommand{\bea}{\begin{eqnarray}}
\newcommand{\eea}{\end{eqnarray}}

\newcommand{\Eq}[1]{Eq.~(\ref{#1})}

\newcommand{\ea}{{\it et al.}}

\newcommand{\np}[1]{{ Nucl. Phys. }{\bf #1}}
\newcommand{\pl}[1]{{ Phys. Lett. }{\bf #1}}

\begin{document}
\thispagestyle{empty}
\begin{flushright} 
UCRHEP-T301\\ 
February 2001\
\end{flushright}
\vspace{0.5in}
\begin{center}
{\Large \bf Neutrino Mass, Muon Anomalous Magnetic Moment,\\and Lepton 
Flavor Nonconservation\\}
\vspace{1.5in}
{\bf Ernest Ma$^1$ and Martti Raidal$^{1,2}$\\}
\vspace{0.2in}
{$^1$ \sl Physics Department, University of California, Riverside, 
California 92521, USA\\}
\vspace{0.1in}
{$^2$ \sl National Institute of Chemical Physics and Biophysics, Tallinn 
10143, Estonia\\}
\vspace{1.5in}
\end{center}
\begin{abstract}\
If the generating mechanism for neutrino mass is to account for both the 
newly observed muon anomalous magnetic moment as well as the present 
experimental bounds on lepton flavor nonconservation, then the neutrino 
mass matrix should be almost degenerate and the underlying physics be 
observable at future colliders.  We illustrate this assertion in 
two specific examples, and show that $\Gamma (\mu \to e \gamma)/m_\mu^5$, 
$\Gamma (\tau \to e \gamma)/m_\tau^5$, and $\Gamma (\tau \to \mu \gamma)
/m_\tau^5$ are in the ratio $(\Delta m^2)_{sol}^2/2$, $(\Delta m^2)_{sol}^2
/2$, and $(\Delta m^2)_{atm}^2$ respectively, where the $\Delta m^2$ 
parameters are those of solar and atmospheric neutrino oscillations and 
bimaximal mixing has been assumed.
\end{abstract}
\newpage
\baselineskip 24pt

Any mechanism for generating a mass matrix for the three neutrinos $\nu_e$, 
$\nu_\mu$, and $\nu_\tau$ will have side effects \cite{whepp6}, among which 
are lepton-flavor violating processes such as $\mu \to e \gamma$, $\tau \to 
\mu \gamma$, and $\mu - e$ conversion in nuclei, as well as an extra 
contribution to the muon anomalous magnetic moment \cite{bnl}.  If the scale 
of this new physics is very high, as in the simplest models of neutrino mass 
\cite{seesaw,masa}, then these side effects are suppressed by the high scale 
and are totally negligible phenomenologically.  However, if this scale is of 
order 1 TeV or less, as in two recent proposals \cite{marasa,ma}, then the 
exciting possibility exists for all of these effects to be visible in 
present and future laboratory experiments.

In view of the newly announced measurement \cite{bnl} of the muon anomalous 
magnetic moment:
\begin{equation}
a_\mu^{exp} = {g_\mu - 2 \over 2} = 116592020 (160) \times 10^{-11},
\label{amu}
\end{equation}
which differs from the standard-model (SM) prediction \cite{czma} by 
$2.6\sigma$:
\begin{equation}
\Delta a_\mu = a_\mu^{exp} - a_\mu^{SM} = 426 \pm 165 \times 10^{-11},
\label{adev}
\end{equation}
a relatively large positive new contribution to $a_\mu$ is needed, hinting 
thus at possible new physics just above the electroweak scale.  One may be 
tempted to believe that it is due to some new physics which has not appeared 
anywhere else before.  On the other hand, a much better established hint of 
new physics already exists, i.e. neutrino mass from neutrino oscillations, 
so it is important to ask the question: {\it Are they related} ?

In this paper we assume that the generating mechanism for neutrino mass is  
responsible for at least a significant part of the deviation shown 
in \Eq{adev}.  We show that unless the neutrino mass matrix is almost 
degenerate, i.e. with 3 nearly equal mass eigenvalues, the $a_\mu$ 
measurement is in conflict with the $\tau \to \mu \gamma$ rate. This is 
because of the nearly maximal $\nu_\mu - \nu_\tau$ mixing for atmospheric 
neutrino oscillations \cite{atm}, as explained below.  We study two 
examples, one of which will be shown to be completely consistent with 
all other flavor-nonconserving processes as well.  We predict the relative 
decay rates of $\mu \to e \gamma,$ $\tau \to e \gamma,$ and $\tau \to \mu 
\gamma$ in terms of neutrino oscillation data, and show that these processes 
constrain the common neutrino mass scale and the solar neutrino oscillation 
solution in a very interesting range.  In addition, the underlying new physics
should be observable at future collider experiments.

Consider the following mass eigenstates of the 3 active neutrinos:
\begin{eqnarray}
\nu_1 &=& \cos \theta \nu_e - {\sin \theta \over \sqrt 2} (\nu_\mu - \nu_\tau),
\\ \nu_2 &=& \sin \theta \nu_e + {\cos \theta \over \sqrt 2} (\nu_\mu - 
\nu_\tau), \\ \nu_3 &=& {1 \over \sqrt 2} (\nu_\mu + \nu_\tau),
\end{eqnarray}
with masses $m_1 \leq m_2 \leq m_3$ respectively.  This choice is dictated 
by the present knowledge of neutrino data regarding atmospheric \cite{atm}
and solar \cite{sol} neutrino oscillations.  
Specifically, $\nu_\mu - \nu_\tau$ mixing is 
assumed to be maximal to explain the atmospheric data  
(we comment on the effect of small allowed deviations from this 
assumption later),
and $\nu_e$ mixes 
with the other two neutrinos with angle $\theta$ to account for the solar 
data. The $3 \times 3$ Majorana neutrino mass matrix in the 
$(\nu_e, \nu_\mu, \nu_\tau)$ basis is then given by
\begin{equation}
{\cal M}_\nu = \left[ \begin{array} {c@{\quad}c@{\quad}c} c^2 m_1 + s^2 m_2 & 
sc(m_2 - m_1)/\sqrt 2 & sc(m_1 - m_2)/\sqrt 2 \\ sc(m_2 - m_1)/\sqrt 2 & 
(s^2 m_1 + c^2 m_2 + m_3)/2 & (-s^2 m_1 - c^2 m_2 + m_3)/2 \\ sc(m_1 - m_2)/
\sqrt 2 & (-s^2 m_1 - c^2m_2 + m_3)/2 & (s^2 m_1 + c^2 m_2 + m_3)/2 
\end{array} \right],
\end{equation}
where $s \equiv \sin \theta$ and $c \equiv \cos \theta$.  For $\theta = 
\pi/4$, it is known as bimaximal mixing.

In the Higgs triplet model \cite{marasa} with $\xi\sim (3,1)$ under the 
standard $SU(2)_L\times U(1)_Y$ gauge group, we have the interaction
\begin{equation}
f_{ij} [\xi^0 \nu_i \nu_j + \xi^+ (\nu_i l_j + l_i \nu_j)/\sqrt 2 + \xi^{++} 
l_i l_j] + h.c.
\label{xiint}
\end{equation}
which gives $({\cal M}_\nu)_{ij} = 2 f_{ij} \langle \xi^0 \rangle,$ 
and establishes a one-to-one correspondence between the neutrino mass 
matrix and the interaction terms. The smallness of ${\cal M}_\nu$ follows 
from the smallness of $\langle \xi^0 \rangle$ \cite{marasa}, while the 
couplings $f_{ij}$ can be large and the triplet mass $m_\xi$ can be the 
order of the electroweak scale.  Therefore, it follows from \Eq{xiint} that 
the muon $g-2$ contribution is proportional to $f^2_{\mu e} + f^2_{\mu \mu} 
+ f^2_{\mu \tau}$, whereas the $\tau \to \mu \gamma$ amplitude is 
proportional to $f_{\tau e} f_{e \mu} + f_{\tau \mu} f_{\mu \mu} + f_{\tau 
\tau} f_{\tau \mu}$.  The former is proportional to $(m_3^2 + c^2 m_2^2 + 
s^2 m_1^2)/2$ and the latter to $(m_3^2 - c^2 m_2^2 - s^2 m_1^2)/2$.  This 
means that a suppression of the $\tau \to \mu \gamma$ rate (relative to the 
muon $g-2$) is possible only if $m_1 \simeq m_2 \simeq m_3$, i.e. a nearly 
degenerate neutrino mass matrix.

In the leptonic Higgs doublet model \cite{ma}, ${\cal M}_\nu$ comes from
the terms
\begin{equation}
{1 \over 2} M_i N^2_{iR} + h_{ij} \bar N_{iR} (\nu_j \eta^0 - l_{jL} \eta^+) 
+ h.c.,
\label{etaint}
\end{equation}
where $\eta\sim(2,1/2)$ and carries lepton number $L=-1,$ while the singlet
fermions $N_R$ have $L=0.$ 
We assume now that all the heavy $N_R$'s are equal in mass.
Hence Eqs.~(3) to (6) imply
\begin{equation}
h_{ij} = \left[ \begin{array} {c@{\quad}c@{\quad}c} 2 c h_1 & -\sqrt 2 s h_1 & 
\sqrt 2 s h_1 \\ 2 s h_2 & \sqrt 2 c h_2 & -\sqrt 2 c h_2 \\ 0 & \sqrt 2 h_3 & 
\sqrt 2 h_3 \end{array} \right],
\end{equation}
with $m_i = 4 h_i^2 \langle \eta^0 \rangle^2/M$.  Again, $m_i$ is small 
because $\langle \eta^0 \rangle$ is small \cite{ma}, allowing thus $h_i$ to 
be large and $M$ the order of the electroweak scale. In this case, the muon 
$g-2$ contribution is proportional to $(m_3 + c^2 m_2 + s^2 m_1)/2$ 
and the $\tau \to \mu \gamma$ amplitude to $(m_3 - c^2 m_2 - s^2 m_1)/2$, 
again suppressing the latter relative to the former in the limit of 
degenerate neutrino masses.

In both of the above models, there are large contributions to 
$\Delta a_\mu$ as 
well as $l_i \to l_j \gamma$ coming from the interactions of Eqs.~(7) and 
(8), as shown in Fig.~1.  In the triplet model,
\bea
\Delta a_\mu=\sum_l \frac{10}{3} {f_{\mu l}^2 \over (4\pi)^2} {m_\mu^2 
\over m_\xi^2}.
\eea
In the limit of a degenerate neutrino mass 
matrix, i.e. $m_1 = m_2 = m_3 = 2 f \langle \xi^0 \rangle$, this implies
\bea
m_\xi < 1174 \sqrt {\alpha_f} ~{\rm GeV},
\label{xibound}
\eea
where $\alpha_f = f^2/4\pi$ and the 90\% confidence-level limit $\Delta a_\mu 
> 215 \times 10^{-11}$ has been used \cite{czma}.  In the doublet model,
\bea
\Delta a_\mu=\sum_i  {h_{i \mu}^2 \over (4\pi)^2} {m_\mu^2 \over m_\eta^2} 
F_2(s_{N_i}),
\eea
where $s_{N_i} \equiv m^2_{N_i}/m^2_\eta$ and
\bea
F_2(x)=\frac{1-6x+3x^2+2x^3-6x^2\ln x}{6(1-x)^4}.
\eea
Assuming $s_{N_i} = 1$ [which gives $F_2(1)=1/12$] and using Eq.~(9) with 
all $h$'s equal, we then obtain
\bea
m_\eta < 371 \sqrt {\alpha_h} ~{\rm GeV},
\label{etabound}
\eea
where $\alpha_h = h^2/4\pi$.  
Comparison of \Eq{xibound} and \Eq{etabound} implies that
masses below 1 TeV are expected in either model.

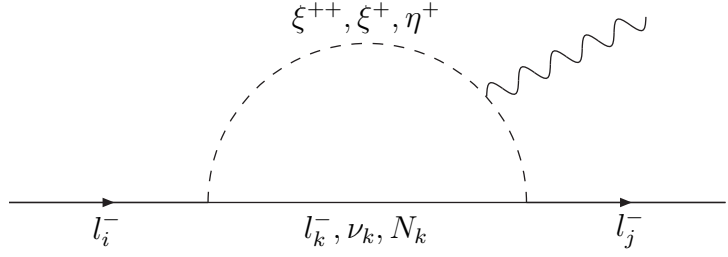
\begin{figure}
\begin{center}
\begin{picture}(270,110)(0,0)
\ArrowLine(0,10)(75,10)
\ArrowLine(195,10)(270,10)
\Line(0,10)(270,10)
\DashCArc(135,10)(60,0,180){4}
\Photon(180,50)(240,80){4}{5}
\Text(135,0)[]{$l^-_k,\nu_k,N_k$}
\Text(37,0)[]{$ l^-_{i}$}
\Text(235,0)[]{$ l^-_{j}$}
\Text(135,80)[]{$\xi^{++},\xi^{+},\eta^+$}
\end{picture}
\end{center}
\caption{Diagrams giving rise to $\Delta a_\mu$ and $l_i \to l_j \gamma$. The 
photon can be attached to any charged line.}
\label{diag}
\end{figure}

The $l_i \to l_j \gamma$ rate divided by the $l_i \to l_j \nu_i \bar \nu_j$ 
rate is given by
\bea
R(l_i\to l_j\gamma)=\frac{96 \pi^3\alpha}{G_F^2m_{l_i}^4}
\left( |f_{M1}|^2 + |f_{E1}|^2 \right) \,,
\eea
where $\alpha \simeq 1/137$ and $G_F$ is the Fermi constant.  In the doublet 
model, the magnetic and electric dipole moment form factors are given by
\bea
f_{M1} = f_{E1}= \sum_k {h_{k l_i} h_{k l_j} \over 4(4\pi)^2} {m_{l_i}^2 
\over m_\eta^2} F_2(1).
\eea
For $\tau \to \mu \gamma$,
\bea
\sum_k h_{k \tau} h_{k \mu} = 2 (h_3^2 - c^2 h_2^2 - s^2 h_1^2) \simeq 
2 (h_3^2 - h_2^2) \simeq {h^2 \over m_\nu^2} (\Delta m^2)_{atm},
\eea
where $m_\nu$ is the common mass of the 3 neutrinos. Hence the $\tau \to \mu 
\gamma$ branching fraction is given by
\bea
B(\tau\to \mu\gamma)= B(\tau\to \mu\nu\bar\nu)
\frac{ \pi\alpha}{192G_F^2} \left( \frac{\alpha_h}{m^2_{\eta}} \right)^2
\frac{(\Delta m^2)^2_{atm}}{m_\nu^4}.
\eea
Suppose we do not have neutrino mass degeneracy, but rather a hierarchical 
neutrino mass matrix, then $(\Delta m^2)_{atm}/m_\nu^2$ would be equal to 
one, and using Eq.~(14), we would obtain $B(\tau \to \mu \gamma) > 8.0 
\times 10^{-6}$, well above the experimental upper limit of $1.1 \times 
10^{-6}$.  Note that this result, while presented for a specific model, 
is actually very general.  If $\nu_3 = c \nu_\mu + s \nu_\tau$, there 
would be a suppression factor of $s^2/c^2$, but this is not available 
because atmospheric neutrino data require nearly maximal $\nu_\mu - \nu_\tau$ 
mixing.

Similarly, the $\mu \to e \gamma$ and $\tau \to e \gamma$ branching fractions 
are given by
\bea
B(\mu\to e\gamma) &=& \frac{ \pi\alpha}{192G_F^2} \left( \frac{\alpha_h}
{m^2_{\eta}} \right)^2 \left[ 2 s^2 c^2 \right]\frac{(\Delta m^2)^2_{sol}}
{m_\nu^4}, \\ B(\tau \to e \gamma) &=& B(\tau \to e \nu \bar \nu) 
B(\mu \to e \gamma).
\eea
Hence we have the interesting relationship
\bea
&& {\Gamma(\mu \to e \gamma) \over m_\mu^5} ~:~ {\Gamma(\tau \to e \gamma) 
\over m_\tau^5} ~:~ {\Gamma(\tau \to \mu \gamma) \over m_\tau^5} \nonumber 
\\ &=& 2s^2c^2 (\Delta m^2)_{sol}^2 ~:~ 2s^2c^2 (\Delta m^2)_{sol}^2 ~:~ 
(\Delta m^2)_{atm}^2.
\eea

The $\mu-e$ conversion ratio $R_{\mu e}$ in nuclei is given by
\bea
R_{\mu e} = {8 \alpha^5 m_\mu^5 Z_{eff}^4 Z |\overline {F_p} (p_e)|^2 \over 
\Gamma_{capt} q^4} \left[|f_{E0} + f_{M1}|^2 + |f_{E1} + f_{M0}|^2 \right],
\eea
where $q^2 \simeq -m_\mu^2$ and for $^{13} Al$, $Z_{eff} = 11.62$, $\overline 
{F_p} = 0.66$, and $\Gamma_{capt} = 7.1 \times 10^5$ s$^{-1}$ \cite{rs,conv}.
The charge-radius form factors are given by
\bea
f_{E0} = -f_{M0}= \sum_i  {h_{i\mu} h_{ie} \over 2(4\pi)^2} {m_\mu^2 \over 
m_\eta^2} F_1(s_{N_i}),
\eea
where
\bea
F_1(x)=\frac{2-9x+18x^2-11x^3+6x^3\ln x}{36(1-x)^4},
\eea
with $F_1(1) = 1/24$.  In Fig.~2, using
\bea
(\Delta m^2)_{atm} = 3 \times 10^{-3} ~{\rm eV}^2,
\eea
and assuming the large-angle matter-enhanced solution of solar neutrino 
oscillations with
\bea
(\Delta m^2)_{sol} = 3 \times 10^{-5} ~{\rm eV}^2,
\eea
we plot $B(\tau \to \mu \gamma)$, $B(\mu \to e \gamma)$, and $R_{\mu e}$ 
as functions of $m_\nu$ for $s^2 = c^2 = 1/2$ and $\alpha_h/m_\eta^2 = (371 
~{\rm GeV})^{-2}$.  Hence these lines should be considered as {\it lower} 
bounds in the case of bimaximal mixing for neutrino oscillations.  

\begin{figure}[t]
\centerline{
\epsfxsize = 0.8\textwidth \epsffile{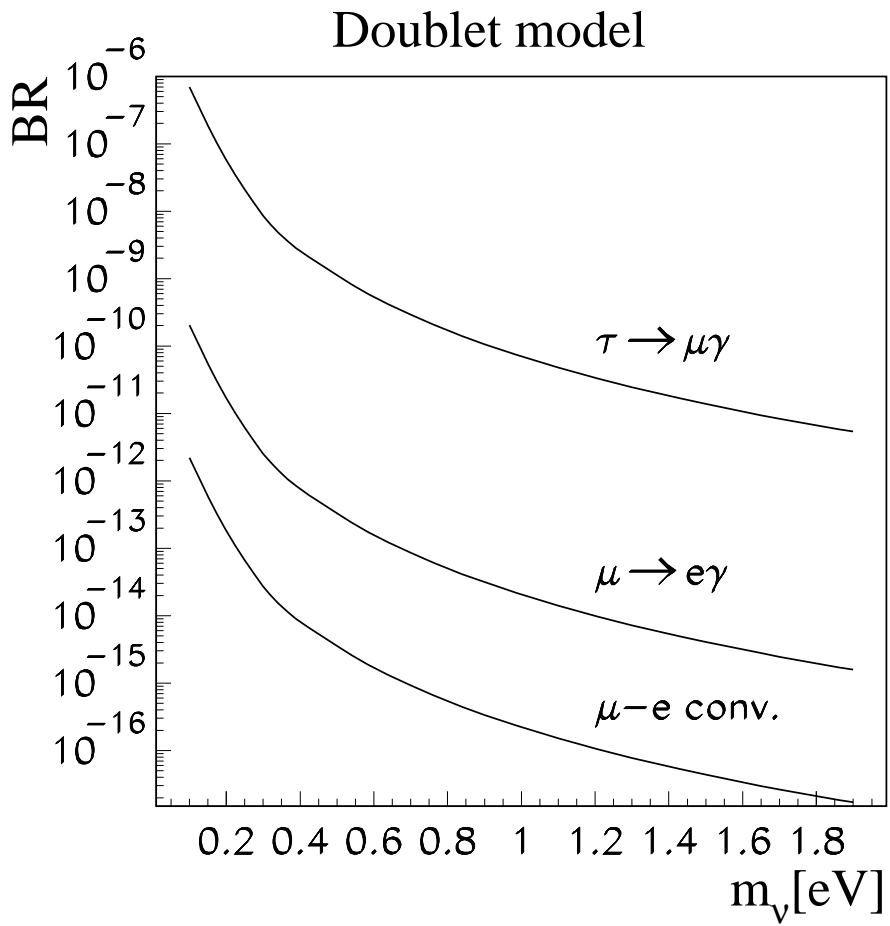} 
}
\caption{Lower bounds on $B(\tau\to \mu \gamma)$, $B(\mu\to e \gamma)$, 
and $R_{\mu e}$ from the measurement of $a_\mu$ in the leptonic Higgs 
doublet model, assuming bimaximal mixing of degenerate neutrinos.}
\label{pl}
\end{figure}

We note 
that at $m_\nu = 0.2$ eV, $B(\mu \to e \gamma)$ is at its present upper limit 
\cite{mueg} of $1.2 \times 10^{-11}$.  If $m_\nu > 0.2$ eV is desired, then 
the constraint from the nonobservation of neutrinoless double beta decay 
\cite{bb} requires the $m_{ee}$ element of Eq.~(6) to be less than 0.2 eV.  
This is easily achieved by making $m_1 < 0$ but keeping $m_{2,3} > 0$, 
without affecting any of our results presented so far.  However, we {\it must} 
then choose the large-angle mixing solution of solar neutrino oscillations, 
implying the observation of $\mu \to e \gamma$ and $\mu-e$ 
Bconversion in the planned experiments with the sensitivities down to $2 
\times 10^{-14}$ \cite{psi} and $2 \times 10^{-17}$ \cite{meco} respectively.  
From Fig.~2 we see that an order-of-magnitude improvement of the present 
$\tau \to \mu \gamma$ bound will also test this specific prediction. 
Thus $B(\tau \to \mu \gamma),$ neutrinoless double beta decay, $B(\mu \to 
e \gamma)$, and $\mu-e$ conversion are all complementary to one another in 
probing the connection between $m_\nu$ and $\Delta a_\mu$.

However, the neutrino mixings need not be exactly bimaximal.
Indeed, the mixing element $|V_{e3}|$ is constrained to be small
but may still be nonzero. Obviously the rate $B(\tau \to \mu \gamma)$ is 
completely independent of this parameter and our conclusion that neutrinos
must be degenerate in mass to explain the observed $ \Delta a_\mu$ remains 
unchanged.
However, $B(\mu \to e \gamma)$, and $R_{\mu e}$ receive additional 
contributions proportional to $|V_{e3}|^2(\Delta m^2)^2_{atm}$ \cite{new}.
For example, if $|V_{e3}|\sim 0.1$ one needs $m_\nu\sim 1$ eV to satisfy
the present experimental bounds. Therefore, no fine tuning in the 
parameters of Fig.~2 is needed to comply with data if $|V_{e3}|\neq 0.$
Nevertheless, the planned $\mu \to e \gamma$ experiments offer sensitive
probes of the small mixing angle $|V_{e3}|$ in this scenario.

In the triplet model, the relevant form factors are calculated in 
Ref. \cite{rs}. We have again the relationship given by Eq.~(21), but 
the corresponding $R_{\mu e}$ is not suppressed as in the doublet model. 
The reason is that the form factors $f_{E0,M0}$ are now functions of 
$m_{l_i}^2/m_\xi^2$ which are
different for different charged leptons, unlike $m_{N_i}^2/m_\eta^2$ which 
are the same for all $N$'s.  As a result, $R_{\mu e}$ is of order $10^{-6}$ 
independent of $m_\nu$, which is definitely ruled out by experiment.  In 
addition, the $\mu \to e e e$ branching fraction [which occurs at tree level, 
but is suppressed by $(\Delta m^2)_{sol}^2$] also exceeds the present 
experimental bound for $m_\nu < 2.7$ eV if $s^2 = c^2 = 1/2$, again assuming 
the large-angle matter-enhanced solution of solar neutrino oscillations.  
Thus the triplet model cannot explain $\Delta a_\mu$ even if neutrino masses 
are degenerate.  It is of course still perfectly viable as a model of neutrino 
masses \cite{marasa}, but it will have no significant contribution to the 
muon $g-2$.

Since the $g-2$ announcement \cite{bnl}, there have been many papers 
\cite{others} dealing with its possible explanation.  Ours is the only one 
relating it to another {\it existing} hint of new physics, i.e. neutrino mass 
from neutrino oscillations.  A glance at Fig.~2 shows that $m_\nu = 0.2$ eV 
is a very interesting number.  It is the present upper limit of a Majorana 
neutrino mass from neutrinoless double beta decay; it also corresponds to the 
present upper limits of $B(\mu \to e \gamma)$ and $\mu-e$ conversion in 
nuclei. Planned experiments on all three fronts are in progress and will 
test our proposed connection between $m_\nu$ and $\Delta a_\mu$.  They 
will also probe the possibly nonzero neutrino mixing angle $V_{e3}$.  
In addition, the $\tau \to \mu \gamma$ branching fraction is just an 
order-of-magnitude away, and Eq.~(14) implies that the leptonic Higgs doublet 
$(\eta^+, \eta^0)$ as well as the fermion singlets $N_{iR}$ are not far away 
from being discovered in future colliders as proposed in Ref.~[6].  A 
neutrino mass of 0.2 eV is also very relevant in cosmology \cite{fulisu} 
and astrophysics \cite{weiler}.

This work was supported in part by the U.~S.~Department of Energy
under Grant No.~DE-FG03-94ER40837 and Estonian ETF grant No. 3832.

\newpage
\bibliographystyle{unsrt}

\newpage
\begin{center}
{\Large \bf Erratum\\}
\end{center}
\vspace{0.2in}
Equation (12) is missing a minus sign on the right-hand side. This renders 
our model as it stands unsuitable for explaining the positive $\Delta a_\mu$ 
observed.  On the other hand, if we extend our model to include supersymmetry 
as recently proposed \cite{nulq}, then a positive $\Delta a_\mu$ from the 
exchange of the supersymmetric particles $\tilde N_i$ and $\tilde \eta$ can 
be obtained.  Equation (12) becomes correct with $m_\eta$ replaced by 
$m_{\tilde \eta}$ and $s_{N_i} \equiv m^2_{\tilde N_i}/m^2_{\tilde \eta}$. 
With this replacement in the rest of our paper, our results remain unchanged.
\bibliographystyle{unsrt}

\end{document}